\documentclass[aps,prd,amssymb,amsmath]{revtex4}

\begin{document}

\title{Stability of semiclassical gravity solutions with respect to
quantum metric fluctuations}
\author{B.~L. Hu}
\author{Albert Roura}
\affiliation{Department of Physics, University of Maryland,
College Park, Maryland 20742-4111}
\author{Enric Verdaguer}
\affiliation{Departament de F\'{\i}sica Fonamental and CER en
Astrof\'\i sica, F\'\i sica de Part\'\i cules i Cosmologia,
Universitat de Barcelona, Av.~Diagonal 647, 08028 Barcelona,
Spain}

\begin{abstract}
We discuss the stability of semiclassical gravity solutions with
respect to small quantum corrections by considering the quantum
fluctuations of the metric perturbations around the semiclassical
solution. We call the attention to the role played by the
symmetrized two-point quantum correlation function for the metric
perturbations,
which can be naturally decomposed into two separate contributions:
intrinsic and induced fluctuations. We show that traditional
criteria on the stability of semiclassical gravity are incomplete
because these criteria based on the linearized semiclassical
Einstein equation can only provide information on the expectation
value and the intrinsic fluctuations of the metric perturbations.
By contrast, the framework of stochastic semiclassical gravity
provides a more complete and accurate criterion because it
contains information on the induced fluctuations as well.
The Einstein-Langevin equation therein contains a
stochastic source characterized by the noise kernel (the symmetrized
two-point quantum correlation function of the stress tensor operator)
and yields stochastic correlation functions for
the metric perturbations which agree, to leading order in the
large $N$ limit, with the quantum correlation functions of the
theory of gravity interacting with $N$ matter fields. These
points are illustrated with the example of Minkowski spacetime as
a solution to the semiclassical Einstein equation, which is found
to be stable under both intrinsic and induced fluctuations.
\end{abstract}


\maketitle

\section{Introduction}
\label{sec1}

In this paper we discuss the stability of the solutions of
semiclassical gravity (SCG) \cite{zeldovich72,hu78,fischetti79,hartle79,birrell94,wald94,flanagan96}
emphasizing the role of
metric fluctuations induced by the quantum matter sources. SCG is
based on the self-consistent solutions of the semiclassical
Einstein equation for a classical spacetime driven by the
expectation value of the stress tensor operator of quantum matter
fields. We propose a criterion based on stochastic semiclassical
gravity which involves the fluctuations of the metric.

SCG accounts for the averaged back reaction of quantum matter
fields and can be regarded as a mean field approximation that
describes the dynamics of the mean spacetime geometry. However,
it does not account for the effects of the fluctuations of
spacetime geometry. Here we focus on the effects of the quantum
fluctuations of the metric. We will restrict our treatment to
small metric perturbations around a given background geometry.
One can then use the stochastic semiclassical gravity formalism
\cite{hu03a,hu04a} to study the fluctuations of the metric
perturbations. In fact, one can show that the leading order
contribution to the quantum correlation functions in a large $N$
expansion is equivalent to the stochastic
correlation functions obtained by solving the Einstein-Langevin
equation in the context of stochastic semiclassical gravity.
By \textit{leading order} in the large $N$
limit we mean the lowest order in $1/N$ with a nonvanishing
contribution
(thus, when
using the rescaled gravitational coupling constant introduced in
Sec.~\ref{sec2.2}, the leading order for the source of the semiclassical
Einstein equation is $1/N^0$, whereas the leading order for the quantum
two-point correlation functions is $1/N$).

Making use of the equivalence between quantum and stochastic
correlation functions in stochastic semiclassical gravity, one is
naturally led to separate the symmetrized quantum correlation
function for the metric perturbations (to leading order in $1/N$)
into two separate contributions: the \emph{intrinsic} and the
\emph{induced} fluctuations. The former is connected to the
dispersion of the initial state of the metric perturbations,
whereas the latter is induced by the quantum fluctuations of the
matter fields' stress tensor operator.

Different aspects concerning the validity of the description
provided by SCG in the case of free quantum matter fields in the
Minkowski vacuum state propagating on Minkowski spacetime have
been studied by a number of authors. Most of them considered the
stability of such a solution of SCG with respect to small
perturbations of the metric. Horowitz was the first one to
analyze the equations describing those perturbations, which
involve higher order derivatives (up to fourth order), and found
unstable solutions that grow exponentially with characteristic
timescales comparable to the Planck time
\cite{horowitz80,horowitz81}. This was later reanalyzed by Jordan
with similar conclusions \cite{jordan87a}. However, those unstable
solutions were regarded as an unphysical artifact by Simon, who
argued that they lie beyond the expected domain of validity of
the theory and emphasized that only those solutions which
resulted from truncating the perturbative expansions in terms of
the square of the Planck length are acceptable
\cite{simon90,simon91}. Further discussion was provided by
Flanagan and Wald \cite{flanagan96}, who advocated the use of an
order reduction prescription first introduced by Parker and Simon
\cite{parker93} but insisted that even nonperturbative solutions
of the resulting second order equation should be regarded as
acceptable. Following these approaches Minkowski metric is shown
to be a stable solution of SCG with respect to small metric
perturbations.

Anderson, Molina-Par\'\i s and Mottola have recently taken up the
issue of the validity of SCG \cite{anderson03} again. Their starting
point is the fact that the semiclassical Einstein equation will fail
to provide a valid description of the dynamics of the mean spacetime
geometry whenever the higher order radiative corrections to the
effective action, involving loops of gravitons or internal graviton
propagators, become important (see
Refs.~\cite{tsamis96a,tsamis96b,tsamis97,tsamis98} for some attempts
to include those effects).  Next, they argue qualitatively that such
higher order radiative corrections cannot be neglected if the metric
fluctuations grow without bound. Finally, they propose a criterion (a
necessary condition) to characterize the growth of the metric
fluctuations, and hence the validity of SCG, based on the stability of
the solutions of the linearized semiclassical equation.

This is a summary of a recent paper we wrote \cite{hu04b}
addressing the issue of the stability of semiclassical solutions
with respect to small quantum corrections. When the metric
perturbations are quantized, the semiclassical equation can be
interpreted as the equation governing the evolution of the
expectation value of the operator for the metric perturbations.
We introduce a stability criterion based on whether the metric
fluctuations grow without bound or not by considering the
behavior of the quantum correlation functions of the metric
perturbations. We emphasize that one should consider not only the
intrinsic fluctuations, but also the induced ones. It is true
that the effect of intrinsic fluctuations can be deduced from an
analysis of the solutions of the perturbed semiclassical Einstein
equation, but in general one cannot retrieve the effect of the
induced fluctuations from it. This effect can be properly
accounted for in the stochastic semiclassical gravity framework.
Both intrinsic and induced fluctuations are innate in the
Einstein-Langevin equation.

Throughout the paper we use natural units with $\hbar=c=1$ and
the $(+,+,+)$ convention of Ref.~\cite{misner73}. We also make use
of the abstract index notation of Ref.~\cite{wald84}. Latin
indices denote abstract indices, whereas Greek indices are
employed when a particular coordinate system is considered.

\section{Semiclassical gravity and stochastic semiclassical gravity}
\label{sec2}

\subsection{Semiclassical gravity}
\label{sec2.1}

A possible first step when addressing the interplay between
gravity and quantum field theory is to consider the evolution of
quantum matter fields (matter field is referred to here as any
field other than the gravitational one) on a classical spacetime
with a nontrivial geometry, characterized by a metric $g_{ab}$.
As opposed to the situation for a Minkowski spacetime, there is
in general no preferred vacuum state for the fields and particle
creation effects naturally arise, such as Hawking radiation for
black holes, cosmological particle creation and the generation of
primordial inhomogeneities in  inflationary cosmological models.
\textbf{Quantum field theory in curved spacetime} (QFTCST) is by
now a well-established subject (at least for free fields and
globally hyperbolic spacetimes) \cite{birrell94,wald94}.

QFTCST is only an approximation in that the matter fields are
treated as test fields evolving on a given spacetime. Einstein's
theory requires that spacetime dynamics determines and is
determined by the matter field. Thus one  needs to consider the
back reaction of the quantum matter fields on the dynamics of the
spacetime geometry, which naturally leads to the
\textbf{semiclassical theory of gravity}, where the evolution of
the spacetime metric $g_{ab}$ is determined by the semiclassical
Einstein equation
\begin{equation}
G_{ab}[g] + \Lambda g_{ab} - \alpha A_{ab}[g] - \beta B_{ab}[g]
=\kappa \left\langle \hat{T}_{ab}[g] \right\rangle' _\mathrm{ren}
, \label{einstein1}
\end{equation}
where $g_{ab}$ is the spacetime metric, $G_{ab}[g]$ is the
Einstein tensor and the matter source corresponds to the
renormalized expectation value of the stress tensor operator of
the matter fields (a prime was used to distinguish it from that
introduced below after absorbing some terms). Here, $\Lambda$ is
the renormalized cosmological constant, $\kappa=8\pi G$, with $G
\equiv 1/m_p^2$ being the Newton constant and $m_p$ the Planck
mass; $\alpha$ and $\beta$ are renormalized dimensionless
coupling constants associated with tensors $A_{ab}[g], B_{ab}[g]$
needed for the renormalization of the logarithmic divergences
(the renormalized coupling constants are running
coupling constants which depend on some renormalization scale
$\mu$; however, since $\langle \hat{T}_{ab}[g] \rangle'
_\mathrm{ren}$ has the same dependence on $\mu$, the
semiclassical Einstein equation is invariant under the
renormalization group, which involves changes in the
renormalization scale $\mu$). The expectation value of the
stress tensor operator exhibits divergences which are local and
state independent.  Introducing a covariant regularization and
renormalization procedure, those divergences can be absorbed into
the cosmological constant, the Newton constant multiplying the
Einstein-Hilbert term and the gravitational action counterterms
quadratic in the curvature.
The finite contributions from those
counterterms give rise to the covariantly conserved tensors
$A_{ab}$ and $B_{ab}$ which result from functionally
differentiating with respect to the metric the terms $\int d^4x
\sqrt{-g}C^{abcd}C_{abcd}$ and $\int d^4x \sqrt{-g} R^2$
respectively, where $C_{abcd}$ is the Weyl tensor and $R$ is the
Ricci scalar. Those contributions were explicitly written on the
left-hand side of Eq.~(\ref{einstein1}), but from now on will be
included in the renormalized expectation value of the stress
tensor operator so that the semiclassical Einstein equation
becomes
\begin{equation}
G_{ab}[g] =\kappa \left\langle \hat{T}_{ab}[g] \right\rangle
_\mathrm{ren} . \label{einstein2}
\end{equation}
The field operators appearing in the stress tensor operator for
the quantum matter fields are in the Heisenberg picture and
satisfy the corresponding equation of motion, which coincides
with the classical field equation for fields evolving on that
spacetime. In particular, if we consider a free scalar field, the
field operator in the Heisenberg picture will satisfy the
corresponding Klein-Gordon equation for that geometry.

Given a manifold $\mathcal{M}$ and a metric $g_{ab}$ which
characterize a globally hyperbolic spacetime, and a density matrix
$\hat{\rho}$ which specifies the state of the quantum matter
fields on a particular Cauchy hypersurface, the triplet
$(\mathcal{M},g_{ab},\hat{\rho})$ constitutes a solution of SCG
if it is a self-consistent solution of both the semiclassical
Einstein equation~(\ref{einstein2}) and the equations of motion
for the quantum operators of the matter fields evolving on the
spacetime manifold $\mathcal{M}$ with metric $g_{ab}$. Those
operators enter in turn into the definition of the stress tensor
operator appearing in the semiclassical Einstein equation.

One can always consider small metric perturbations around a given
solution of semiclassical gravity characterized by a metric $g_{ab}$.
The linearized semiclassical equation for the metric perturbations
becomes then
\begin{equation}
G_{ab}^{(1)}\left[ g+h\right] = \kappa \left\langle
\hat{T}_{ab}^{(1)} [g+h] \right\rangle _\mathrm{ren}
\label{einstein4},
\end{equation}
where the superindex $(1)$ was used to denote that only terms linear
in the metric perturbation $h_{ab}$ should be considered.  The
expectation value $\langle \hat{T}_{ab}^{(1)} [g+h] \rangle
_\mathrm{ren}$ can be evaluated working directly with the quantum
operators for the matter fields in the Heisenberg picture in some cases
\cite{martin99b}, but is usually more convenient to obtain it from the
corresponding effective action in the CTP formalism
\cite{jordan86,calzetta87,campos94}.

\subsection{Stochastic semiclassical gravity}
\label{sec2.2}

The semiclassical Einstein equation, which takes into account only
the mean values, is inadequate whenever the fluctuations of the
stress tensor operator are important. An improved treatment is
provided by the Einstein-Langevin equation of \textbf{stochastic
gravity}, which contains a (Gaussian) stochastic source with a
vanishing expectation value and a correlation function
characterized by the symmetrized two-point function of the stress
tensor operator. This theory has been discussed by a number of
authors
\cite{calzetta94,hu95a,hu95b,campos96,calzetta97c,martin99b,hu03a,hu04a}.
Consider a globally hyperbolic background spacetime and an
initial state for the quantum matter fields (one usually
restricts to free fields) which constitute a self-consistent solution of
SCG, \emph{i.e.}, they satisfy the semiclassical Einstein
equation with the expectation value of the stress tensor operator
obtained by considering the evolution of the matter fields on the
same background geometry. The Einstein-Langevin equation
governing the dynamics of the linearized perturbations $h_{ab}$
around the background metric $g_{ab}$ is given by
\begin{equation}
G_{ab}^{(1)}\left[ g+h\right] =\kappa \left\langle
\hat{T}_{ab}^{(1)} [g+h] \right\rangle _\mathrm{ren}+\kappa \,\xi
_{ab}\left[ g\right] , \label{langevin}
\end{equation}
where the Gaussian stochastic source $\xi _{ab}[g]$ is completely
characterized by its correlation function in terms of the noise
kernel $\mathcal{N}_{abcd}(x,y)$, which accounts for the
fluctuations of the stress tensor operator, as follows:
\begin{equation}
\left\langle \xi _{ab}[g;x) \xi _{cd}[g;y) \right\rangle _{\xi } =
\mathcal{N}_{abcd}(x,y) \equiv \frac{1}{2}\left\langle
\left\{\hat{t}_{ab}[g;x), \hat{t}_{cd}[g;y) \right\}
\right\rangle , \label{noise}
\end{equation}
where $\hat{t}_{ab} \equiv
\hat{T}_{ab}-\langle\hat{T}_{ab}\rangle$ and $\langle \ldots
\rangle$ is the usual expectation value with respect to the
quantum state of the matter fields, whereas $\langle \ldots
\rangle_\xi$ denotes taking the average with respect to all
possible realizations of the stochastic source $\xi_{ab}$. Note
that any local term quadratic in the curvature arising from
finite contributions of the counterterms required to renormalize
the bare expectation value of the stress tensor operator has been
absorbed into its renormalized version $\langle \hat{T}_{ab}^{(1)}
[g+h] \rangle_\mathrm{ren}$.  It should also be emphasized that
solutions of the Einstein-Langevin equation for the metric
perturbations are classical \emph{stochastic} tensorial fields,
not quantum operators.

The precise meaning that should be given to these stochastic metric
perturbations and the relation of the corresponding stochastic
correlation functions to the quantum fluctuations that result from
quantizing these metric perturbations can be established by
considering $N$ matter fields.  Making use of a large $N$ expansion,
one can then show that the stochastic correlation functions for the
metric perturbations obtained from the Einstein-Langevin equation
coincide with the leading order contribution to the quantum
correlation functions in the large $N$ limit \cite{hu04b,roura03b}.
In particular, the two-point stochastic correlation function is
equivalent to the symmetrized quantum correlation function to leading
order in $1/N$ provided that one also averages over the initial
conditions for the solutions of the Einstein-Langevin equation
distributed according to the Wigner functional characterizing the
initial state of the metric perturbations (see Eq.~(C11) in
Ref.~\cite{hu04b} for the definition of the Wigner functional). It is,
therefore, convenient to express the solutions of the
Einstein-Langevin equation as
\begin{equation}
h_{ab} (x) = \Sigma_{ab}^{(0)}(x) + \bar{\kappa} (G_\mathrm{ret}
\cdot \xi)_{ab} (x) \label{solution2},
\end{equation}
where we have introduced the notacion 
$A\cdot B \equiv \int d^4 y \sqrt{-g(y)}A(y)B(y)$,
$\bar{\kappa} = N \kappa$ is the rescaled gravitational
coupling constant introduced in Ref.~\cite{hu04b},
$\Sigma_{ab}^{(0)}(x)$ is a solution of the homogeneous part of
the Einstein-Langevin equation~(\ref{langevin}) containing all the
information about the initial conditions (by homogeneous part we
mean Eq.~(\ref{langevin}) excluding the stochastic source, which
coincides with the semiclassical Einstein
equation~(\ref{einstein2})), and $G_\mathrm{ret} (x,x')$ is the
retarded propagator with vanishing initial conditions associated
with that equation (see Appendix~E~3 in Ref.~\cite{hu04b}
for important remarks on the propagator). Using
Eq.~(\ref{noise}), we can then get the following result for the
symmetrized two-point quantum correlation function of the metric
perturbations around a Minkowski background:
\begin{equation}
\frac{1}{2}\left\langle \left\{ \hat{h}_{ab}(x), \hat{h}_{cd}(x')
\right\} \right\rangle = \left\langle \Sigma_{ab}^{(0)}(x)
\Sigma_{cd}^{(0)}(x') \right\rangle _{\Sigma_{ab}^{(i)},
\Pi^{cd}_{(i)}} + \frac{\bar{\kappa}^2}{N} \left( G_\mathrm{ret}
\cdot \mathcal{N} \cdot (G_\mathrm{ret})^{T} \right)_{abcd}
(x,x') \label{correlation2},
\end{equation}
where the Lorentz gauge condition $\nabla^a ({h}_{ab} - (1/2)
\eta_{ab} h^c_c) = 0$ as well as some initial condition to fix
completely the remaining gauge freedom of the initial state
should be implicitly understood, and the stochastic source was
rescaled according to Refs.~\cite{hu04b,roura03b} so that $\langle \xi
_{ab}[g;x) \xi _{cd}[g;y) \rangle_{\xi } = (1/N)
\mathcal{N}_{abcd}(x,y)$, where $\mathcal{N}_{abcd}(x,y)$ is the
noise kernel for a single field.

There are two different contributions to
the symmetrized quantum correlation function. The first one is
connected to the quantum fluctuations of the initial state of the
metric perturbations and we will refer to it as \emph{intrinsic
fluctuations}. The second contribution, proportional to the noise
kernel, accounts for the fluctuations due to the interaction with
the matter fields, and we will refer to it as \emph{induced
fluctuations}.

\section{Stability of semiclassical gravity solutions: previous work}
\label{sec3}

Although the stability of other semiclassical gravity solutions in
addition to Minkowski spacetime has been studied (see, for instance,
Refs.~\cite{starobinsky80,vilenkin85,suen89} for analysis involving
Robertson-Walker geometries), most of the analysis have concentrated
on the stability of small perturbations around Minkowski spacetime.
This case already exhibits the main features and difficulties that one
may encounter when dealing with back-reaction effects in semiclassical
gravity and will be used in the next Section to illustrate the
generalized stability criterion introduced there. In this
Section  we give a brief review of previous work on the stability of
semiclassical gravity solutions specialized, for the reasons mentioned
above, to the case of Minkowski spacetime.

The stability of metric perturbations around a Minkowski spacetime
interacting with quantum matter fields in their Minkowski vacuum state
was first studied in the context of SCG by Horowitz \cite{horowitz80}.
He considered massless conformally coupled scalar fields and found
exponential instabilities for the linearized metric perturbations with
characteristic timescales comparable to the Planck time. Those
solutions are closely related to the higher derivative countertems
required to renormalize the expectation value of the stress tensor
operator and are analogous to the runaway solutions commonly present
in radiation reaction processes such as those considered in classical
electrodynamics \cite{jackson99,johnson02}. It is generally believed
that the runaway solutions obtained by Horowitz are an unphysical
artifact since they involve scales beyond the regime where SCG is
expected to be reliable (in fact, this statement can be naturally
formulated when regarding general relativity as a low energy effective
theory \cite{burgess03}).

Since the existence of terms with higher derivatives in time
implies an increase in the number of degrees of freedom (in an
initial value formulation, not only the metric and its time
derivative should be specified, but also its second and third
order time derivatives), it seems plausible that, by restricting
to an appropriate subspace of solutions of the semiclassical
Einstein equation, one can reestablish the usual number of
degrees of freedom in general relativity and, at the same time,
get rid of all the unphysical runaway solutions. Following this
line of thought Simon proposed that one should restrict to
solutions which result from truncating to order $\hbar$ an
analytic expansion in $\hbar$ (or equivalently in $l_p^2$, the
Planck length squared) \cite{simon90,simon91}. Together with
Parker he also introduced a prescription to reduce the order of
the semiclassical Einstein equation which was computationally
convenient in order to obtain solutions corresponding to such
truncated perturbative expansions in $\hbar$ \cite{parker93}.

On the other hand, Flanagan and Wald argued that Simon's criterion
based on truncating to order $\hbar$ solutions which correspond to
analytic expansions in $\hbar$ seemed too restrictive since it
only allowed small deviations with respect to the classical
solutions of the Einstein equations \cite{flanagan96}. In
particular, one would miss those situations in which the small
semiclassical corrections build up to give significant deviations
at long times, such as those corresponding to the evaporation of
a macroscopic black hole (with a mass much larger than the Planck
mass) by emission of Hawking radiation. Furthermore, they
illustrated with simple examples that there are cases in which
one expects that no solutions of the semiclassical equation are
analytic in $\hbar$. Therefore, they suggested that, rather than
trying to restrict the subspace of acceptable solutions, one
should simply transform the semiclassical equation, by making use
of Simon and Parker's order reduction prescription, to a second
order equation which were equivalent to the original equation up
to the order in $\hbar$ (or $l_p^2$) under consideration. All the
solutions of the second order equation should then be regarded as
acceptable, even if they are not analytic in $\hbar$. Obviously,
one could only extract physically reliable information from those
solutions for scales much larger than the Planck length.

Yet another prescription was proposed by  Anderson,
Molina-Par\'\i s and Mottola \cite{anderson03} on the stability
of small metric perturbations around the Minkowski spacetime.
They got rid of the unphysical runaway solutions by working in
Fourier space and discarding those solutions which corresponded
to 4-momenta with modulus comparable or larger in absolute value
than the Planck mass. However, it is not clear how this procedure
could be generalized to situations where working in Fourier space
is not adequate, as in  time-dependent background spacetimes.

The consequences of both the order reduction prescription
introduced by Simon and Parker and advocated by Flanagan and
Wald, and the procedure employed by Anderson \emph{et al.} are
rather drastic, at least when applied to the case of a Minkowski
background, since one is just left with the solutions of the
sourceless classical Einstein equation corresponding to linear
gravitational waves propagating in Minkowski. In fact, the
situation was not completely trivial for Flanagan and Wald, who
were interested in analyzing whether the averaged null energy
condition (ANEC) was satisfied in SCG by considering
perturbations of the Minkowski solution, because they also
perturbed the state of the matter fields. The order reduction
prescription also seems to exclude those solutions which
correspond to inflationary models driven entirely by the vacuum
polarization of the quantum matter fields \cite{simon92}, such as
the trace anomaly driven inflationary model initially proposed by
Starobinsky \cite{starobinsky80}. To keep this kind of models,
Hawking, Hertog and Reall considered a less drastic alternative
to deal with the runaway solutions \cite{hawking01,hawking02}.
Their procedure, which is analogous to some methods previously
employed in classical electrodynamics for radiation reaction
problems \cite{jackson99}, is based on discarding solutions which
grow without bound at late times (see Ref.~\cite{hu04b} for
further discussions on this and related issues).

\section{Generalized stability criterion. Application to Minkowski
spacetime}
\label{sec4}

\subsection{Generalized stability criterion}
\label{sec4.1}

How does one characterize the quantum state of the metric
perturbations? The first candidate is the expectation value of the
operator associated with the perturbation of the metric,
$\hat{h}_{ab}$. In fact, using a large $N$ expansion, Hartle and
Horowitz showed that the semiclassical Einstein equation can be
interpreted as the equation governing the evolution of the expectation
value of the metric to leading order in $1/N$ \cite{hartle81}.  Taking
that result into account, the study of the stability of a solution of
SCG by linearizing the semiclassical Einstein equation with respect to
small metric perturbations around that solution can be understood in
the following way: Take an initial state for the metric perturbations
with a small nonvanishing expectation value for the operator
$\hat{h}_{ab}$, let it evolve, and see if the expectation value grows
without bound.

However, in addition to the expectation value of $\hat{h}_{ab}$ the
state of the metric perturbations will also be characterized by its
fluctuations. Let us now suppose that the evolution of the expectation
value is stable (\emph{i.e.}, that it does not grow unboundedly with
time) or even that it vanishes for all times. It is clear that the
semiclassical solution cannot be regarded as stable with respect to
small quantum corrections if the fluctuations of the state for the
metric perturbations grow without bound. Therefore, the stability
criteria based on the solutions of the semiclassical Einstein
equation, which can be interpreted as conditions on the stability of
the expectation value of the operator $\hat{h}_{ab}$ for the state of
the metric perturbations, should be generalized: one also needs to
take into account the fluctuations. In addition to the expectation
value, the $n$-point quantum correlation functions for the metric
perturbations (starting with $n=2$) should also be stable.

As explained in Refs.~\cite{hu04b,roura03b}, to leading order in $1/N$
the CTP generating functional for the metric perturbations exhibits a
Gaussian form provided that a Gaussian initial state for the metric
perturbations with vanishing expectation value is chosen. All the
$n$-point quantum correlation functions can then be obtained, to
leading order in $1/N$, from the two-point quantum correlation
function. Furthermore, any of the two-point quantum correlation
functions can in turn be expressed in terms of the symmetrized and
antisymmetrized correlation functions (the expectation values of the
commutator and anticommutator of the operator $\hat{h}_{ab}$). To
leading order in $1/N$ the commutator is independent of the initial
state of the metric perturbations and is given by $2i\kappa
(G_\mathrm{ret}(x',x) - G_\mathrm{ret}(x,x'))$. On the other hand, the
expectation value of the anticommutator is given by
Eq.~(\ref{correlation2}) and is the sum of two separate contributions:
the intrinsic and the induced fluctuations.

The first contribution in Eq.~(\ref{correlation2}) to the correlation
function for the metric perturbations involves the solutions of the
homogeneous part of the Einstein-Langevin equation~(\ref{langevin}),
which actually coincides with the linearized semiclassical equation for
the metric perturbations around the background geometry. Similarly,
$G_\mathrm{ret}$ corresponds to the retarded propagator (with
vanishing initial conditions) associated with the linearized semiclassical
equation. Thus, solving the perturbed semiclassical Einstein equation
not only accounts for the evolution of the expectation value of the metric
perturbations, which will exhibit a nontrivial dynamics as long as we
choose an initial state with nonvanishing expectation value, but also
provides nontrivial information, even for a state with a vanishing
expectation value, about the commutator as well as the intrinsic
fluctuations of the metric. This implies that the analysis about the
stability of the solutions of SCG can also be used to
determine the stability of the metric perturbations with respect to
intrinsic fluctuations.

The observation we make here is that the induced fluctuations can
be important as well. Both the retarded propagator and the solutions
of the linearized semiclassical Einstein equation depend
on the expectation value of the commutator of the stress
tensor operator on the background geometry and on the imaginary part
of its time-ordered two-point function. However, they do not involve
the expectation value of the anticommutator, which drives the induced
fluctuations. Furthermore, although the expectation value of the
commutator and the anticommutator are related by a
fluctuation-dissipation relation in some particular cases
\cite{martin99b,martin00}, that is not true in general and the induced
fluctuations need to be explicitly analyzed.

To sum up, when analyzing the stability of a solution of SCG
with respect to small quantum corrections, one should also
consider the behavior of both the intrinsic and induced
fluctuations of the quantized metric perturbations. Whereas
information on the stability of the intrinsic fluctuations can be
retrieved from an analysis of the solutions of the perturbed
semiclassical Einstein equation, the effect of the induced
fluctuations is properly accounted for only in the stochastic
semiclassical gravity framework based on the Einstein-Langevin
equation.

\subsection{Stability of Minkowski space from our criterion}
\label{sec4.2}

We now turn to the application of the criterion proposed in the
previous subsection to the particular yet important case of
Minkowski spacetime. As explained there, the existing results in
the literature can be interpreted as analysis of the stability of
the expectation value of the operator associated with the metric
perturbations (see, however,
Refs.~\cite{hartle81,horowitz81,jaekel95}). On the other hand, we
also need to include in our consideration the fluctuations,
characterized by the two-point quantum correlation function.

In order to analyze the two-point quantum correlation function for
the metric perturbations, we will exploit the fact that the
stochastic correlation functions obtained with the solutions of the
Einstein-Langevin equation coincide with the quantum correlation
functions for the metric perturbations. Moreover, according to
Eq.~(\ref{correlation2}), the symmetrized two-point quantum
correlation function has two different contributions: the intrinsic
and the induced fluctuations. We proceed now to analyze each
contribution separately.

The first term on the right-hand side of Eq.~(\ref{correlation2})
corresponds to the fluctuations of the metric perturbations due to
the fluctuations of their initial state and is given by
\begin{equation}
\left\langle \Sigma_{ab}^{(0)}(x) \Sigma_{cd}^{(0)}(x') \right\rangle
_{\Sigma_{ab}^{(i)}, \Pi^{cd}_{(i)}}
\label{intrinsic},
\end{equation}
where we recall that $\Sigma_{ab}^{(0)}(x)$ is a solution of
the homogeneous part of the Einstein-Langevin equation (once the
Lorentz gauge has been imposed) with the appropriate initial
conditions.

As mentioned in Sec.~\ref{sec2.2}, the homogeneous part of the
Einstein-Langevin equation actually coincides with the linearized
semiclassical Einstein equation~(\ref{einstein4}). Therefore, we can
make use of the results derived in
Refs.~\cite{horowitz80,flanagan96,anderson03}, which are briefly
summarized in Appendix~E of Ref.~\cite{hu04b}. As described there, in
addition to the solutions with $G_{ab}^{(1)}(x)=0$, there are other
solutions that in Fourier space take the form
$\tilde{G}^{(1)}_{\mu\nu}(p) \propto \delta(p^2-p_0^2)$ for some
particular values of $p_0^2$, but they all exhibit exponential
instabilities with Planckian characteristic timescales.

In order to deal with those unstable solutions, one possibility
is to employ the order reduction prescription. We are then left
only with the solutions which satisfy $\tilde{G}_{\mu\nu}^{(1)}(p)=0$
(see Ref.~\cite{hu04b}). The result for the metric
perturbations in the gauge introduced above can be obtained by
solving for the Einstein tensor in that gauge:
$\tilde{G}^{(1)}_{ab}(p) = (1/2) p^2 (\tilde{h}_{\mu\nu} (p) -
1/2 \eta_{\mu\nu} \tilde{h}^\rho_\rho (p))$. Those solutions for
$\tilde{h}_{\mu\nu}(p)$ simply correspond to free linear
gravitational waves propagating in Minkowski spacetime expressed
in the transverse and traceless (TT) gauge. When substituting back
into Eq.~(\ref{intrinsic}) and averaging over the initial
conditions we simply get the symmetrized quantum correlation
function for free gravitons in the TT gauge for the state given
by the reduced Wigner function.

A second possibility, proposed by Hawking \emph{et al.}
\cite{hawking01,hawking02}, is to impose boundary conditions which
discard the runaway solutions that grow unboundedly in time and
correspond to a special prescription for the integration contour when
Fourier transforming back to spacetime coordinates (see Appendix~E in
Ref.~\cite{hu04b} for a more detailed discussion). Following that
procedure we get, for example, that for a massless conformally coupled
scalar field, with appropriate values of the renormalized coupling
constants, the intrinsic contribution to the symmetrized quantum
correlation function coincides with that of free gravitons plus an
extra contribution for the scalar part of the metric perturbations
which renders Minkowski spacetime stable but plays a crucial
role in providing a graceful exit for inflationary models driven by
the vacuum polarization of a large number of conformal fields (such a
massive scalar field would not be in conflict with present
observations because, for the range of parameters usually considered,
the mass would be far too large to have observational consequences
\cite{hawking01}).

The second term on the right-hand side of Eq.~(\ref{correlation2})
corresponds to the fluctuations of the metric perturbations induced
by the fluctuations of the quantum matter fields and is given by
\begin{equation}
\frac{\bar{\kappa}^2}{N} \left( G_\mathrm{ret} \cdot \mathcal{N}
\cdot (G_\mathrm{ret})^{T} \right)_{abcd} (x,x')
= N \kappa^2 \left( G_\mathrm{ret} \cdot \mathcal{N}
\cdot (G_\mathrm{ret})^{T} \right)_{abcd} (x,x')
\label{induced},
\end{equation}
where $\mathcal{N}_{abcd}(x,x')$ is the noise kernel accounting for
the fluctuations of the stress tensor operator, and
$(G_\mathrm{ret})_{abcd}(x,x')$ is the retarded propagator with
vanishing initial conditions associated with the integro-differential
operator $L_{abcd}(x,x')$ defined as
\begin{equation}
L_{abcd}(x,x') = (1/2) (\eta_{ac}\eta_{bd} - \eta_{ab}\eta_{cd}/2)
\Box \delta(x-x') + 2 \bar{\kappa} H^\mathrm{(ren)}_{abcd}(x-x')
+ 2 \bar{\kappa} M^\mathrm{(ren)}_{abcd}(x-x') \label{linearop},
\end{equation}
where the kernel $H$ corresponds to the sum of the expectation values
of the commutator and the imaginary part of the time-ordered product
of the stress tensor operator for the matter fields evaluated on the
background geometry, and the kernel $M$ is obtained by functionally
differentiating with respect to the metric the expectation value of
the stress tensor operator on the background geometry taking into
account only its explicit dependence on the metric. See Eqs.~(5) and
(6) in Ref.~\cite{hu04b} for the exact definition of both kernels.

The same kind of exponential instabilities as in the runaway solutions of
the homogeneous part of the Einstein-Langevin equation (the linearized
semiclassical Einstein equation) also arise when computing the
retarded propagator $G_\mathrm{ret}$. In order to deal with those
instabilities, similar to the case of the intrinsic fluctuations, one
possibility is to make use of the order reduction prescription. The
Einstein-Langevin equation becomes then $G_{ab}^{(1)} = \kappa
\xi_{ab}$. The second possibility, following the proposal of Hawking
\emph{et al.}, is to impose boundary conditions which discard the
exponentially growing solutions and translate into a special choice of
the integration contour when Fourier transforming back to spacetime
coordinates the expression for the propagator. In fact, it turns out
that the propagator which results from adopting that prescription
coincides with the propagator that was employed in
Ref.~\cite{martin00}. However, it should be emphasized that this
propagator is no longer the retarded one since it exhibits causality
violations at Planckian scales. A more detailed discussion on all
these points can be found in Appendix~E of Ref.~\cite{hu04b}.

Following Refs.~\cite{martin00,hu04b}, the Einstein-Langevin equation
can be entirely written in terms of the linearized Einstein tensor
$\tilde{G}^{(1)}_{\mu\nu}(p)$. One can then solve the stochastic
equation for $\tilde{G}^{(1)}_{\mu\nu}(p)$ and obtain its correlation
function \cite{martin00,hu04b}:
\begin{eqnarray}
\langle \tilde{G}^{(1)}_{\mu\nu}(p) \tilde{G}^{(1)}_{\rho\sigma}(p')
\rangle_\xi &=& \bar{\kappa}^2 \tilde{D}_{\mu\nu\alpha\beta}(p)
\langle \tilde{\xi}^{\alpha\beta}(p) \tilde{\xi}^{\gamma\delta}(p')
\rangle_\xi \tilde{D}_{\rho\sigma\gamma\delta}(p') \nonumber \\
&=& \frac{\bar{\kappa}^2}{N} \tilde{D}_{\mu\nu\alpha\beta}(p)
\tilde{\mathcal{N}}^{\alpha\beta\gamma\delta}(p)
\tilde{D}_{\rho\sigma\gamma\delta}(-p) (2\pi)^4 \delta(p+p')
\label{fourierGG1},
\end{eqnarray}
where the explicit expressions for the noise kernel
$\tilde{\mathcal{N}}^{\alpha\beta\gamma\delta}(p)$ and the propagator
$\tilde{D}_{\mu\nu\alpha\beta}(p)$ can be found, respectively, in
Appendices~B and E of Ref.~\cite{hu04b}. On the other hand, if we
make use of the order reduction prescription, we get
\begin{equation}
\langle \tilde{G}^{(1)}_{\mu\nu}(p) \tilde{G}^{(1)}_{\rho\sigma}(p')
\rangle_\xi = \bar{\kappa}^2 \langle \tilde{\xi}^{\alpha\beta}(p)
\tilde{\xi}^{\gamma\delta}(p') \rangle_\xi = \frac{\bar{\kappa}^2}{N}
\tilde{\mathcal{N}}^{\alpha\beta\gamma\delta}(p) (2\pi)^4 \delta(p+p')
\label{fourierGG2}.
\end{equation}
Note that $G^{(1)}_{\mu\nu}(p)$ is gauge invariant when
perturbing a Minkowski background because the background tensor
$G^{(0)}_{ab}$ vanishes and, hence, $\mathcal{L}_{\vec{\zeta}}
G^{(0)}_{ab}$ also vanishes for any vector field $\vec{\zeta}$.

Finally, using the expression for the linearized Einstein tensor in
the Lorentz gauge, $\tilde{G}^{(1)}_{\mu\nu} = (1/2) p^2
\tilde{\bar{h}}_{\mu\nu}$ with $\bar{h}_{\mu\nu} = h_{\mu\nu}
- (1/2) \eta_{\mu\nu} h^{\alpha}_{\alpha}$, we obtain the correlation
function for the metric perturbations in that gauge:
\begin{equation}
\langle \tilde{\bar{h}}_{\mu\nu}(p) \tilde{\bar{h}}_{\rho\sigma}(p')
\rangle_\xi = \frac{4\bar{\kappa}^2}{N} \frac{1}{(p^2)^2}
\tilde{D}_{\mu\nu\alpha\beta}(p)
\tilde{\mathcal{N}}^{\alpha\beta\gamma\delta}(p)
\tilde{D}_{\rho\sigma\gamma\delta}(-p) (2\pi)^4 \delta(p+p')
\label{fourierhh1},
\end{equation}
or
\begin{equation}
\langle \tilde{\bar{h}}_{\mu\nu}(p) \tilde{\bar{h}}_{\rho\sigma}(p')
\rangle_\xi = \frac{4\bar{\kappa}^2}{N} \frac{1}{(p^2)^2}
\tilde{\mathcal{N}}_{\mu\nu\rho\sigma}(p) (2\pi)^4 \delta(p+p')
\label{fourierhh2},
\end{equation}
if the order reduction prescription is employed. It should be
emphasized that, contrary to the linearized Einstein tensor
$G^{(1)}_{ab}$, the metric perturbation $h_{ab}$ is not gauge
invariant. This should not pose a major problem provided that the
gauge has been completely fixed, as explained in
Refs.~\cite{roura03b,hu04b}.

The correlation functions in spacetime coordinates can be easily
derived by Fourier transforming Eqs.~(\ref{fourierhh1}) or
(\ref{fourierhh2}). There is apparently an infrared divergence at $p^2
= 0$ for the massless case, but such an infrared divergence seems to
be just a gauge artifact \cite{hu04b}. Therefore, we can conclude
that, once the instabilities giving rise to the unphysical runaway
solutions have been properly dealt with, the fluctuations of the
metric perturbations around the Minkowski spacetime induced by the
interaction with quantum scalar fields are indeed stable (if
instabilities had been present, they would have led to a divergent
result when Fourier transforming back to spacetime coordinates).

\section{Discussion}
\label{sec5}

An analysis of the stability of any solution of SCG with respect
to small quantum corrections should consider not only the
evolution of the expectation value of the metric perturbations
around that solution, but also their fluctuations, encoded in the
quantum correlation functions. Making use of the equivalence (to
leading order in $1/N$, where $N$ is the number of matter fields)
between the stochastic correlation functions obtained in
stochastic semiclassical gravity and the quantum correlation
functions for metric perturbations around a solution of SCG, the
symmetrized two-point quantum correlation function for the metric
perturbations can be decomposed into two distinct parts: the
intrinsic fluctuations due to the fluctuations of the initial
state of the metric perturbations itself, and the fluctuations
induced by their interaction with the matter fields. If one
considers the linearized perturbations of the semiclassical
Einstein equation, only information on the intrinsic fluctuations
can be retrieved. On the other hand, the information on the
induced fluctuations naturally follows from the solutions of the
Einstein-Langevin equation.

As a specific example, we analyzed the symmetrized two-point quantum
correlation function for the metric perturbations around the Minkowski
spacetime interacting with $N$ scalar fields initially in the
Minkowski vacuum state. Once the ultraviolet instabilities which are
ubiquitous in SCG \cite{hu04b} and are commonly regarded as
unphysical, have been properly dealt with by using the order reduction
prescription or the procedure proposed in
Refs.~\cite{hawking01,hawking02}, both the intrinsic and the induced
contributions to the quantum correlation function for the metric
perturbations are found to be stable.

The symmetrized quantum correlation function obtained for the
metric perturbations around Minkowski is in agreement with the
real part of the propagator obtained by Tomboulis in
Ref.~\cite{tomboulis77} using a large $N$ expansion
(he actually
considered fermionic rather than scalar fields, but that just
amounts to a change in one coefficient). It is worth noticing that the
imaginary part of the propagator
can be easily obtained from the expectation value
for the commutator of the metric perturbations, which is given by
$2i\kappa(G_\mathrm{ret} (x',x) - G_\mathrm{ret} (x,x'))$, as
explained in Refs.~\cite{hu04b,roura03b}.
Tomboulis used the
\emph{in-out} formalism rather than the CTP formalism employed in
this paper.  Nevertheless, his propagator is equivalent to the
time-ordered CTP propagator when asymptotic initial conditions
are considered because in Minkowski spacetime there is no real
particle creation and the \emph{in} and \emph{out} vacua are
equivalent (up to some phase which is absorbed in the usual
normalization of the \emph{in-out} propagator). The use of a CTP
formulation is, however, crucial to obtain true correlation
functions rather than transition matrix elements in dynamical
(nonstationary) situations (such as in an expanding
Robertson-Walker background geometry), where the \emph{in-out}
scattering matrix might not even be well defined.

It should be mentioned that Ford and collaborators have stressed the
importance of the metric fluctuations and investigated some of their
physical implications
\cite{ford82,kuo93,ford99,ford03,borgman03,ford97,yu99,yu00}.  They
have considered both intrinsic \cite{ford97,ford99,yu99,yu00} and
induced fluctuations \cite{ford82,kuo93,ford99,ford03,borgman03},
which they usually refer to as \emph{active} and \emph{passive}
fluctuations, respectively. However, they usually consider these two
kinds of fluctuations separately and have not provided a unified
treatment where both of them can be understood as different
contributions to the full quantum correlation function.  Moreover,
they always neglect the nonlocal term which encodes the averaged back
reaction on the metric perturbations due to the modified dynamics of
the matter fields generated by the metric perturbations themselves
(this term is often called the dissipation term by analogy with
quantum Brownian motion models).
Their justification is by arguing that those terms
would be of higher order in a perturbative expansion. That is
indeed the case when considering a Minkowski background if the
order reduction prescription is employed, but it is not clear
whether it remains true under more general conditions. In fact,
as mentioned in Ref.~\cite{roura03a}, for the usual cosmological
inflationary models the contribution of the nonlocal terms can be
comparable or even larger than that of the remaining terms.
Finally, in order to deal with the singular coincidence limit of
the noise kernel, in Ref.~\cite{kuo93} Ford and collaborators
opted to subtract a number of terms including the fluctuations
for the Minkowski vacuum. Even when no such subtraction was
performed (because a method based on multiple integrations by
parts was used instead) \cite{ford99,wu01,wu02}, they usually
discard the fluctuations for the Minkowski vacuum. Therefore, the
information on the metric fluctuations around a Minkowski
background when the matter fields are in the vacuum state is
missing in their work.

An additional number of partially open issues are discussed in
\cite{hu04b},  to which the reader is referred for further
details.

\begin{acknowledgments}
It is a pleasure to thank Enrique \'{A}lvarez, Paul Anderson,
Daniel Arteaga, Dieter Brill, Larry Ford, Carmen
Molina-Par\'{\i}s and Emil Mottola for interesting discussions.
B.~L.~H. and A.~R. are supported by NSF under Grant PHY03-00710.
E.~V. acknowledges support from the MICYT Research Project
No.~FPA-2001-3598.
\end{acknowledgments}


\begin{thebibliography}{57}
\expandafter\ifx\csname natexlab\endcsname\relax\def\natexlab#1{#1}\fi
\expandafter\ifx\csname bibnamefont\endcsname\relax
  \def\bibnamefont#1{#1}\fi
\expandafter\ifx\csname bibfnamefont\endcsname\relax
  \def\bibfnamefont#1{#1}\fi
\expandafter\ifx\csname citenamefont\endcsname\relax
  \def\citenamefont#1{#1}\fi
\expandafter\ifx\csname url\endcsname\relax
  \def\url#1{\texttt{#1}}\fi
\expandafter\ifx\csname urlprefix\endcsname\relax\def\urlprefix{URL }\fi
\providecommand{\bibinfo}[2]{#2}
\providecommand{\eprint}[2][]{\url{#2}}

\bibitem[{\citenamefont{Zeldovich and Starobinsky}(1972)}]{zeldovich72}
\bibinfo{author}{\bibfnamefont{Y.~B.} \bibnamefont{Zeldovich}}
  \bibnamefont{and} \bibinfo{author}{\bibfnamefont{A.~A.}
  \bibnamefont{Starobinsky}}, \bibinfo{journal}{Sov. Phys. JETP}
  \textbf{\bibinfo{volume}{34}}, \bibinfo{pages}{1159} (\bibinfo{year}{1972}).

\bibitem[{\citenamefont{Hu and Parker}(1978)}]{hu78}
\bibinfo{author}{\bibfnamefont{B.~L.} \bibnamefont{Hu}} \bibnamefont{and}
  \bibinfo{author}{\bibfnamefont{L.}~\bibnamefont{Parker}},
  \bibinfo{journal}{Phys. Rev. D} \textbf{\bibinfo{volume}{17}},
  \bibinfo{pages}{933} (\bibinfo{year}{1978}).

\bibitem[{\citenamefont{Fischetti et~al.}(1979)\citenamefont{Fischetti, Hartle,
  and Hu}}]{fischetti79}
\bibinfo{author}{\bibfnamefont{M.~V.} \bibnamefont{Fischetti}},
  \bibinfo{author}{\bibfnamefont{J.~B.} \bibnamefont{Hartle}},
  \bibnamefont{and} \bibinfo{author}{\bibfnamefont{B.~L.} \bibnamefont{Hu}},
  \bibinfo{journal}{Phys. Rev. D} \textbf{\bibinfo{volume}{20}},
  \bibinfo{pages}{1757} (\bibinfo{year}{1979}).

\bibitem[{\citenamefont{Hartle and Hu}(1979)}]{hartle79}
\bibinfo{author}{\bibfnamefont{J.~B.} \bibnamefont{Hartle}} \bibnamefont{and}
  \bibinfo{author}{\bibfnamefont{B.~L.} \bibnamefont{Hu}},
  \bibinfo{journal}{Phys. Rev. D} \textbf{\bibinfo{volume}{20}},
  \bibinfo{pages}{1772} (\bibinfo{year}{1979}).

\bibitem[{\citenamefont{Birrell and Davies}(1994)}]{birrell94}
\bibinfo{author}{\bibfnamefont{N.~D.} \bibnamefont{Birrell}} \bibnamefont{and}
  \bibinfo{author}{\bibfnamefont{P.~C.~W.} \bibnamefont{Davies}},
  \emph{\bibinfo{title}{Quantum fields in curved space}}
  (\bibinfo{publisher}{Cambridge University Press},
  \bibinfo{address}{Cambridge}, \bibinfo{year}{1994}).

\bibitem[{\citenamefont{Wald}(1994)}]{wald94}
\bibinfo{author}{\bibfnamefont{R.~M.} \bibnamefont{Wald}},
  \emph{\bibinfo{title}{Quantum field theory in curved spacetime and black hole
  thermodynamics}} (\bibinfo{publisher}{The University of Chicago Press},
  \bibinfo{address}{Chicago}, \bibinfo{year}{1994}).

\bibitem[{\citenamefont{Flanagan and Wald}(1996)}]{flanagan96}
\bibinfo{author}{\bibfnamefont{E.~E.} \bibnamefont{Flanagan}} \bibnamefont{and}
  \bibinfo{author}{\bibfnamefont{R.~M.} \bibnamefont{Wald}},
  \bibinfo{journal}{Phys. Rev. D} \textbf{\bibinfo{volume}{54}},
  \bibinfo{pages}{6233} (\bibinfo{year}{1996}).

\bibitem[{\citenamefont{Hu and Verdaguer}(2003)}]{hu03a}
\bibinfo{author}{\bibfnamefont{B.~L.} \bibnamefont{Hu}} \bibnamefont{and}
  \bibinfo{author}{\bibfnamefont{E.}~\bibnamefont{Verdaguer}},
  \bibinfo{journal}{Class. Quant. Grav.} \textbf{\bibinfo{volume}{20}},
  \bibinfo{pages}{R1} (\bibinfo{year}{2003}).

\bibitem[{\citenamefont{Hu and Verdaguer}(2004)}]{hu04a}
\bibinfo{author}{\bibfnamefont{B.~L.} \bibnamefont{Hu}} \bibnamefont{and}
  \bibinfo{author}{\bibfnamefont{E.}~\bibnamefont{Verdaguer}},
  \bibinfo{journal}{Living Rev. Rel.} \textbf{\bibinfo{volume}{7}},
  \bibinfo{pages}{3} (\bibinfo{year}{2004}).

\bibitem[{\citenamefont{Horowitz}(1980)}]{horowitz80}
\bibinfo{author}{\bibfnamefont{G.~T.} \bibnamefont{Horowitz}},
  \bibinfo{journal}{Phys. Rev. D} \textbf{\bibinfo{volume}{21}},
  \bibinfo{pages}{1445} (\bibinfo{year}{1980}).

\bibitem[{\citenamefont{Horowitz}(1981)}]{horowitz81}
\bibinfo{author}{\bibfnamefont{G.~T.} \bibnamefont{Horowitz}}, in
  \emph{\bibinfo{booktitle}{Quantum gravity 2: a second {O}xford symposium}},
  edited by \bibinfo{editor}{\bibfnamefont{C.~J.} \bibnamefont{Isham}},
  \bibinfo{editor}{\bibfnamefont{R.}~\bibnamefont{Penrose}}, \bibnamefont{and}
  \bibinfo{editor}{\bibfnamefont{D.~W.} \bibnamefont{Sciama}}
  (\bibinfo{publisher}{Clarendon Press}, \bibinfo{address}{Oxford, United
  Kingdom}, \bibinfo{year}{1981}).

\bibitem[{\citenamefont{Jordan}(1987)}]{jordan87a}
\bibinfo{author}{\bibfnamefont{R.~D.} \bibnamefont{Jordan}},
  \bibinfo{journal}{Phys. Rev. D} \textbf{\bibinfo{volume}{36}},
  \bibinfo{pages}{3593} (\bibinfo{year}{1987}).

\bibitem[{\citenamefont{Simon}(1990)}]{simon90}
\bibinfo{author}{\bibfnamefont{J.~Z.} \bibnamefont{Simon}},
  \bibinfo{journal}{Phys. Rev. D} \textbf{\bibinfo{volume}{41}},
  \bibinfo{pages}{3720} (\bibinfo{year}{1990}).

\bibitem[{\citenamefont{Simon}(1991)}]{simon91}
\bibinfo{author}{\bibfnamefont{J.~Z.} \bibnamefont{Simon}},
  \bibinfo{journal}{Phys. Rev. D} \textbf{\bibinfo{volume}{43}},
  \bibinfo{pages}{3308} (\bibinfo{year}{1991}).

\bibitem[{\citenamefont{Parker and Simon}(1993)}]{parker93}
\bibinfo{author}{\bibfnamefont{L.}~\bibnamefont{Parker}} \bibnamefont{and}
  \bibinfo{author}{\bibfnamefont{J.~Z.} \bibnamefont{Simon}},
  \bibinfo{journal}{Phys. Rev. D} \textbf{\bibinfo{volume}{47}},
  \bibinfo{pages}{1339} (\bibinfo{year}{1993}).

\bibitem[{\citenamefont{Anderson et~al.}(2003)\citenamefont{Anderson,
  Molina-Par\'{\i}s, and Mottola}}]{anderson03}
\bibinfo{author}{\bibfnamefont{P.~R.} \bibnamefont{Anderson}},
  \bibinfo{author}{\bibfnamefont{C.}~\bibnamefont{Molina-Par\'{\i}s}},
  \bibnamefont{and} \bibinfo{author}{\bibfnamefont{E.}~\bibnamefont{Mottola}},
  \bibinfo{journal}{Phys. Rev. D} \textbf{\bibinfo{volume}{67}},
  \bibinfo{pages}{024026} (\bibinfo{year}{2003}).

\bibitem[{\citenamefont{Tsamis and Woodard}(1996{\natexlab{a}})}]{tsamis96a}
\bibinfo{author}{\bibfnamefont{N.~C.} \bibnamefont{Tsamis}} \bibnamefont{and}
  \bibinfo{author}{\bibfnamefont{R.~P.} \bibnamefont{Woodard}},
  \bibinfo{journal}{Nucl. Phys. B} \textbf{\bibinfo{volume}{474}},
  \bibinfo{pages}{235} (\bibinfo{year}{1996}{\natexlab{a}}).

\bibitem[{\citenamefont{Tsamis and Woodard}(1996{\natexlab{b}})}]{tsamis96b}
\bibinfo{author}{\bibfnamefont{N.~C.} \bibnamefont{Tsamis}} \bibnamefont{and}
  \bibinfo{author}{\bibfnamefont{R.~P.} \bibnamefont{Woodard}},
  \bibinfo{journal}{Phys. Rev. D} \textbf{\bibinfo{volume}{54}},
  \bibinfo{pages}{2621} (\bibinfo{year}{1996}{\natexlab{b}}).

\bibitem[{\citenamefont{Tsamis and Woodard}(1997)}]{tsamis97}
\bibinfo{author}{\bibfnamefont{N.~C.} \bibnamefont{Tsamis}} \bibnamefont{and}
  \bibinfo{author}{\bibfnamefont{R.~P.} \bibnamefont{Woodard}},
  \bibinfo{journal}{Ann. Phys. (NY)} \textbf{\bibinfo{volume}{253}},
  \bibinfo{pages}{1} (\bibinfo{year}{1997}).

\bibitem[{\citenamefont{Tsamis and Woodard}(1998)}]{tsamis98}
\bibinfo{author}{\bibfnamefont{N.~C.} \bibnamefont{Tsamis}} \bibnamefont{and}
  \bibinfo{author}{\bibfnamefont{R.~P.} \bibnamefont{Woodard}},
  \bibinfo{journal}{Phys. Rev. D} \textbf{\bibinfo{volume}{57}},
  \bibinfo{pages}{4826} (\bibinfo{year}{1998}).

\bibitem[{\citenamefont{Hu et~al.}(2004)\citenamefont{Hu, Roura, and
  Verdaguer}}]{hu04b}
\bibinfo{author}{\bibfnamefont{B.~L.} \bibnamefont{Hu}},
  \bibinfo{author}{\bibfnamefont{A.}~\bibnamefont{Roura}}, \bibnamefont{and}
  \bibinfo{author}{\bibfnamefont{E.}~\bibnamefont{Verdaguer}},
  \bibinfo{journal}{Phys. Rev. D} \textbf{\bibinfo{volume}{70}},
  \bibinfo{pages}{044002} (\bibinfo{year}{2004}).

\bibitem[{\citenamefont{Misner et~al.}(1973)\citenamefont{Misner, Thorne, and
  Wheeler}}]{misner73}
\bibinfo{author}{\bibfnamefont{C.~W.} \bibnamefont{Misner}},
  \bibinfo{author}{\bibfnamefont{K.~S.} \bibnamefont{Thorne}},
  \bibnamefont{and} \bibinfo{author}{\bibfnamefont{J.~A.}
  \bibnamefont{Wheeler}}, \emph{\bibinfo{title}{Gravitation}}
  (\bibinfo{publisher}{Freeman}, \bibinfo{address}{San Francisco},
  \bibinfo{year}{1973}).

\bibitem[{\citenamefont{Wald}(1984)}]{wald84}
\bibinfo{author}{\bibfnamefont{R.~M.} \bibnamefont{Wald}},
  \emph{\bibinfo{title}{General Relativity}} (\bibinfo{publisher}{The
  University of Chicago Press}, \bibinfo{address}{Chicago},
  \bibinfo{year}{1984}).

\bibitem[{\citenamefont{Mart\'{\i }n and Verdaguer}(1999)}]{martin99b}
\bibinfo{author}{\bibfnamefont{R.}~\bibnamefont{Mart\'{\i }n}}
  \bibnamefont{and}
  \bibinfo{author}{\bibfnamefont{E.}~\bibnamefont{Verdaguer}},
  \bibinfo{journal}{Phys. Rev. D} \textbf{\bibinfo{volume}{60}},
  \bibinfo{pages}{084008} (\bibinfo{year}{1999}).

\bibitem[{\citenamefont{Jordan}(1986)}]{jordan86}
\bibinfo{author}{\bibfnamefont{R.~D.} \bibnamefont{Jordan}},
  \bibinfo{journal}{Phys. Rev. D} \textbf{\bibinfo{volume}{33}},
  \bibinfo{pages}{444} (\bibinfo{year}{1986}).

\bibitem[{\citenamefont{Calzetta and Hu}(1987)}]{calzetta87}
\bibinfo{author}{\bibfnamefont{E.}~\bibnamefont{Calzetta}} \bibnamefont{and}
  \bibinfo{author}{\bibfnamefont{B.~L.} \bibnamefont{Hu}},
  \bibinfo{journal}{Phys. Rev. D} \textbf{\bibinfo{volume}{35}},
  \bibinfo{pages}{495} (\bibinfo{year}{1987}).

\bibitem[{\citenamefont{Campos and Verdaguer}(1994)}]{campos94}
\bibinfo{author}{\bibfnamefont{A.}~\bibnamefont{Campos}} \bibnamefont{and}
  \bibinfo{author}{\bibfnamefont{E.}~\bibnamefont{Verdaguer}},
  \bibinfo{journal}{Phys. Rev. D} \textbf{\bibinfo{volume}{49}},
  \bibinfo{pages}{1861} (\bibinfo{year}{1994}).

\bibitem[{\citenamefont{Calzetta and Hu}(1994)}]{calzetta94}
\bibinfo{author}{\bibfnamefont{E.}~\bibnamefont{Calzetta}} \bibnamefont{and}
  \bibinfo{author}{\bibfnamefont{B.~L.} \bibnamefont{Hu}},
  \bibinfo{journal}{Phys. Rev. D} \textbf{\bibinfo{volume}{49}},
  \bibinfo{pages}{6636} (\bibinfo{year}{1994}).

\bibitem[{\citenamefont{Hu and Matacz}(1995)}]{hu95a}
\bibinfo{author}{\bibfnamefont{B.~L.} \bibnamefont{Hu}} \bibnamefont{and}
  \bibinfo{author}{\bibfnamefont{A.}~\bibnamefont{Matacz}},
  \bibinfo{journal}{Phys. Rev. D} \textbf{\bibinfo{volume}{51}},
  \bibinfo{pages}{1577} (\bibinfo{year}{1995}).

\bibitem[{\citenamefont{Hu and Sinha}(1995)}]{hu95b}
\bibinfo{author}{\bibfnamefont{B.~L.} \bibnamefont{Hu}} \bibnamefont{and}
  \bibinfo{author}{\bibfnamefont{S.}~\bibnamefont{Sinha}},
  \bibinfo{journal}{Phys. Rev. D} \textbf{\bibinfo{volume}{51}},
  \bibinfo{pages}{1587} (\bibinfo{year}{1995}).

\bibitem[{\citenamefont{Campos and Verdaguer}(1996)}]{campos96}
\bibinfo{author}{\bibfnamefont{A.}~\bibnamefont{Campos}} \bibnamefont{and}
  \bibinfo{author}{\bibfnamefont{E.}~\bibnamefont{Verdaguer}},
  \bibinfo{journal}{Phys. Rev. D} \textbf{\bibinfo{volume}{53}},
  \bibinfo{pages}{1927} (\bibinfo{year}{1996}).

\bibitem[{\citenamefont{Calzetta et~al.}(1997)\citenamefont{Calzetta, Campos,
  and Verdaguer}}]{calzetta97c}
\bibinfo{author}{\bibfnamefont{E.}~\bibnamefont{Calzetta}},
  \bibinfo{author}{\bibfnamefont{A.}~\bibnamefont{Campos}}, \bibnamefont{and}
  \bibinfo{author}{\bibfnamefont{E.}~\bibnamefont{Verdaguer}},
  \bibinfo{journal}{Phys. Rev. D} \textbf{\bibinfo{volume}{56}},
  \bibinfo{pages}{2163} (\bibinfo{year}{1997}).

\bibitem[{\citenamefont{Roura and Verdaguer}({\natexlab{a}})}]{roura03b}
\bibinfo{author}{\bibfnamefont{A.}~\bibnamefont{Roura}} \bibnamefont{and}
  \bibinfo{author}{\bibfnamefont{E.}~\bibnamefont{Verdaguer}},
  \bibinfo{note}{in preparation}.

\bibitem[{\citenamefont{Starobinsky}(1980)}]{starobinsky80}
\bibinfo{author}{\bibfnamefont{A.~A.} \bibnamefont{Starobinsky}},
  \bibinfo{journal}{Phys. Lett. B} \textbf{\bibinfo{volume}{91}},
  \bibinfo{pages}{99} (\bibinfo{year}{1980}).

\bibitem[{\citenamefont{Vilenkin}(1985)}]{vilenkin85}
\bibinfo{author}{\bibfnamefont{A.}~\bibnamefont{Vilenkin}},
  \bibinfo{journal}{Phys. Rev. D} \textbf{\bibinfo{volume}{32}},
  \bibinfo{pages}{2511} (\bibinfo{year}{1985}).

\bibitem[{\citenamefont{Suen}(1989)}]{suen89}
\bibinfo{author}{\bibfnamefont{W.~M.} \bibnamefont{Suen}},
  \bibinfo{journal}{Phys. Rev. Lett.} \textbf{\bibinfo{volume}{62}},
  \bibinfo{pages}{2217} (\bibinfo{year}{1989}).

\bibitem[{\citenamefont{Jackson}(1999)}]{jackson99}
\bibinfo{author}{\bibfnamefont{J.~D.} \bibnamefont{Jackson}},
  \emph{\bibinfo{title}{Classical electrodynamics}}
  (\bibinfo{publisher}{Wiley}, \bibinfo{address}{New York},
  \bibinfo{year}{1999}).

\bibitem[{\citenamefont{Johnson and Hu}(2002)}]{johnson02}
\bibinfo{author}{\bibfnamefont{P.~R.} \bibnamefont{Johnson}} \bibnamefont{and}
  \bibinfo{author}{\bibfnamefont{B.~L.} \bibnamefont{Hu}},
  \bibinfo{journal}{Phys. Rev. D} \textbf{\bibinfo{volume}{65}},
  \bibinfo{pages}{065015} (\bibinfo{year}{2002}).

\bibitem[{\citenamefont{Burgess}(2003)}]{burgess03}
\bibinfo{author}{\bibfnamefont{C.~P.} \bibnamefont{Burgess}},
  \bibinfo{journal}{Living Rev. Rel.} \textbf{\bibinfo{volume}{7}},
  \bibinfo{pages}{5} (\bibinfo{year}{2004}).

\bibitem[{\citenamefont{Simon}(1992)}]{simon92}
\bibinfo{author}{\bibfnamefont{J.~Z.} \bibnamefont{Simon}},
  \bibinfo{journal}{Phys. Rev. D} \textbf{\bibinfo{volume}{45}},
  \bibinfo{pages}{1953} (\bibinfo{year}{1992}).

\bibitem[{\citenamefont{Hawking et~al.}(2001)\citenamefont{Hawking, Hertog, and
  Reall}}]{hawking01}
\bibinfo{author}{\bibfnamefont{S.~W.} \bibnamefont{Hawking}},
  \bibinfo{author}{\bibfnamefont{T.}~\bibnamefont{Hertog}}, \bibnamefont{and}
  \bibinfo{author}{\bibfnamefont{H.~S.} \bibnamefont{Reall}},
  \bibinfo{journal}{Phys. Rev. D} \textbf{\bibinfo{volume}{63}},
  \bibinfo{pages}{083504} (\bibinfo{year}{2001}).

\bibitem[{\citenamefont{Hawking and Hertog}(2002)}]{hawking02}
\bibinfo{author}{\bibfnamefont{S.~W.} \bibnamefont{Hawking}} \bibnamefont{and}
  \bibinfo{author}{\bibfnamefont{T.}~\bibnamefont{Hertog}},
  \bibinfo{journal}{Phys. Rev. D} \textbf{\bibinfo{volume}{65}},
  \bibinfo{pages}{103515} (\bibinfo{year}{2002}).

\bibitem[{\citenamefont{Hartle and Horowitz}(1981)}]{hartle81}
\bibinfo{author}{\bibfnamefont{J.~B.} \bibnamefont{Hartle}} \bibnamefont{and}
  \bibinfo{author}{\bibfnamefont{G.~T.} \bibnamefont{Horowitz}},
  \bibinfo{journal}{Phys. Rev. D} \textbf{\bibinfo{volume}{24}},
  \bibinfo{pages}{257} (\bibinfo{year}{1981}).

\bibitem[{\citenamefont{Mart\'{\i }n and Verdaguer}(2000)}]{martin00}
\bibinfo{author}{\bibfnamefont{R.}~\bibnamefont{Mart\'{\i }n}}
  \bibnamefont{and}
  \bibinfo{author}{\bibfnamefont{E.}~\bibnamefont{Verdaguer}},
  \bibinfo{journal}{Phys. Rev. D} \textbf{\bibinfo{volume}{61}},
  \bibinfo{pages}{124024} (\bibinfo{year}{2000}).

\bibitem[{\citenamefont{Jaekel and Reynaud}(1995)}]{jaekel95}
\bibinfo{author}{\bibfnamefont{M.~T.} \bibnamefont{Jaekel}} \bibnamefont{and}
  \bibinfo{author}{\bibfnamefont{S.}~\bibnamefont{Reynaud}},
  \bibinfo{journal}{Ann. Phys. (Leipzig)} \textbf{\bibinfo{volume}{4}},
  \bibinfo{pages}{68} (\bibinfo{year}{1995}).

\bibitem[{\citenamefont{Tomboulis}(1977)}]{tomboulis77}
\bibinfo{author}{\bibfnamefont{E.}~\bibnamefont{Tomboulis}},
  \bibinfo{journal}{Phys. Lett. B} \textbf{\bibinfo{volume}{70}},
  \bibinfo{pages}{361} (\bibinfo{year}{1977}).

\bibitem[{\citenamefont{Ford}(1982)}]{ford82}
\bibinfo{author}{\bibfnamefont{L.~H.} \bibnamefont{Ford}},
  \bibinfo{journal}{Ann. Phys. (NY)} \textbf{\bibinfo{volume}{144}},
  \bibinfo{pages}{238} (\bibinfo{year}{1982}).

\bibitem[{\citenamefont{Kuo and Ford}(1993)}]{kuo93}
\bibinfo{author}{\bibfnamefont{C.~I.} \bibnamefont{Kuo}} \bibnamefont{and}
  \bibinfo{author}{\bibfnamefont{L.~H.} \bibnamefont{Ford}},
  \bibinfo{journal}{Phys. Rev. D} \textbf{\bibinfo{volume}{47}},
  \bibinfo{pages}{4510} (\bibinfo{year}{1993}).

\bibitem[{\citenamefont{Ford}(1999)}]{ford99}
\bibinfo{author}{\bibfnamefont{L.~H.} \bibnamefont{Ford}},
  \bibinfo{journal}{Int. J. Theor. Phys.} \textbf{\bibinfo{volume}{38}},
  \bibinfo{pages}{2941} (\bibinfo{year}{1999}).

\bibitem[{\citenamefont{Ford and Wu}(2003)}]{ford03}
\bibinfo{author}{\bibfnamefont{L.~H.} \bibnamefont{Ford}} \bibnamefont{and}
  \bibinfo{author}{\bibfnamefont{C.~H.} \bibnamefont{Wu}},
  \bibinfo{journal}{Int. J. Theor. Phys.} \textbf{\bibinfo{volume}{42}},
  \bibinfo{pages}{15} (\bibinfo{year}{2003}).

\bibitem[{\citenamefont{Borgman and Ford}(2003)}]{borgman03}
\bibinfo{author}{\bibfnamefont{J.}~\bibnamefont{Borgman}} \bibnamefont{and}
  \bibinfo{author}{\bibfnamefont{L.~H.} \bibnamefont{Ford}}
  (\bibinfo{year}{2003}), \eprint{gr-qc/0307043}.

\bibitem[{\citenamefont{Ford and Svaiter}(1997)}]{ford97}
\bibinfo{author}{\bibfnamefont{L.~H.} \bibnamefont{Ford}} \bibnamefont{and}
  \bibinfo{author}{\bibfnamefont{N.~F.} \bibnamefont{Svaiter}},
  \bibinfo{journal}{Phys. Rev. D} \textbf{\bibinfo{volume}{56}},
  \bibinfo{pages}{2226} (\bibinfo{year}{1997}).

\bibitem[{\citenamefont{Yu and Ford}(1999)}]{yu99}
\bibinfo{author}{\bibfnamefont{H.}~\bibnamefont{Yu}} \bibnamefont{and}
  \bibinfo{author}{\bibfnamefont{L.~H.} \bibnamefont{Ford}},
  \bibinfo{journal}{Phys. Rev. D} \textbf{\bibinfo{volume}{60}},
  \bibinfo{pages}{084023} (\bibinfo{year}{1999}).

\bibitem[{\citenamefont{Yu and Ford}(2000)}]{yu00}
\bibinfo{author}{\bibfnamefont{H.}~\bibnamefont{Yu}} \bibnamefont{and}
  \bibinfo{author}{\bibfnamefont{L.~H.} \bibnamefont{Ford}},
  \bibinfo{journal}{Phys. Lett. B} \textbf{\bibinfo{volume}{496}},
  \bibinfo{pages}{107} (\bibinfo{year}{2000}).

\bibitem[{\citenamefont{Roura and Verdaguer}({\natexlab{b}})}]{roura03a}
\bibinfo{author}{\bibfnamefont{A.}~\bibnamefont{Roura}} \bibnamefont{and}
  \bibinfo{author}{\bibfnamefont{E.}~\bibnamefont{Verdaguer}},
  \bibinfo{note}{in preparation}.

\bibitem[{\citenamefont{Wu and Ford}(2001)}]{wu01}
\bibinfo{author}{\bibfnamefont{C.~H.} \bibnamefont{Wu}} \bibnamefont{and}
  \bibinfo{author}{\bibfnamefont{L.~H.} \bibnamefont{Ford}},
  \bibinfo{journal}{Phys. Rev. D} \textbf{\bibinfo{volume}{64}},
  \bibinfo{pages}{045010} (\bibinfo{year}{2001}).

\bibitem[{\citenamefont{Wu et~al.}(2002)\citenamefont{Wu, Kuo, and
  Ford}}]{wu02}
\bibinfo{author}{\bibfnamefont{C.~H.} \bibnamefont{Wu}},
  \bibinfo{author}{\bibfnamefont{C.~I.} \bibnamefont{Kuo}}, \bibnamefont{and}
  \bibinfo{author}{\bibfnamefont{L.~H.} \bibnamefont{Ford}},
  \bibinfo{journal}{Phys. Rev. A} \textbf{\bibinfo{volume}{65}},
  \bibinfo{pages}{062102} (\bibinfo{year}{2002}).

\end{thebibliography}

\end{document}